\RequirePackage[2020-02-02]{latexrelease}
\documentclass[12pt]{revtex4}

\usepackage{multirow}
\usepackage{amsmath,amsfonts}
\usepackage{amssymb,latexsym}
\usepackage{color}
\usepackage{graphicx}
\usepackage{mathrsfs}

\newtheorem{prop}{Proposition}

\newcommand{\beprop}{\begin{prop}}
\newcommand{\enprop}{\end{prop}}
\newcommand{\bprf}{\begin{proof}} 
\newcommand{\eprf}{\end{proof}\qed}

\definecolor{hervecolor}{rgb}{0.8,0,0.7}

\newcommand{\ket}[1]{|\kern.3ex#1\kern.3ex\rangle}
\newcommand{\bra}[1]{\langle\kern.3ex #1 \kern.3ex|}
\newcommand{\scalar}[2]{\langle\kern.3ex #1 \kern.3ex|\kern.3ex#2\kern.3ex\rangle}

\newcommand{\ii}{\mathsf{i}}

\newcommand{\ud}{\mathrm{d}}

\newcommand{\uE}{\mathrm{E}}
\newcommand{\uD}{\mathrm{D}}

\begin{document}

\title{Functional Ideal Hydrodynamics incorporating Quantum-Field Theoretical Fluctuation}
\author{T.\ Koide}
\email{tomoikoide@gmail.com,koide@if.ufrj.br}
\affiliation{Instituto de F\'{\i}sica, Universidade Federal do Rio de Janeiro, C.P. 68528,
21941-972, Rio de Janeiro, RJ, Brazil}
\author{T. Kodama}
\email{kodama.takeshi@gmail.com,koide@if.ufrj.br}
\affiliation{Instituto de F\'{\i}sica, Universidade Federal do Rio de Janeiro, C.P. 68528,
21941-972, Rio de Janeiro, RJ, Brazil}
\affiliation{Instituto de F\'{\i}sica, Universidade Federal Fluminense, 24210-346,
Niter\'{o}i, RJ, Brazil}
\begin{abstract}
We propose new ideal hydrodynamics in the function space which describes a fluid composed of the 1+1 dimensional real scalar field 
in the framework of the stochastic variational method (SVM).
In the derivation, the thermal equilibrium is assumed to the internal state of fluid elements in the function space of the scalar-field configuration.
The deterministic trajectory of the functional fluid element is related to the functional generalization of the Bohmian trajectory in relativistic quantum field theory.
To find the correspondence relation to standard hydrodynamics, a further coarse-graining should be introduced.
Thus functional hydrodynamics is regarded as a mesoscopic theory such as the Boltzmann equation 
in the dynamical hierarchy of many-body systems.
Functional hydrodynamics reproduces the exact behaviors of relativistic quantum field theory in a certain limit.
We thus expect that our theory is applicable to study the influence of quantum-field theoretical fluctuation in collective flows 
of produced particles in relativistic heavy-ion collisions. 
\end{abstract}

\maketitle

\section{introduction}

Hydrodynamics has been used as an effective model to study many-body effects and collective motions in not only classical but also quantum systems.
For example, the QCD vacuum is excited by relativistic heavy-ion collisions and its collective behavior is well described by relativistic hydrodynamical models \cite{hydro_review}. 
In the formulation of hydrodynamics, a fluid is often divided into coarse-grained particles called the fluid elements.
Each fluid element has a constant of motion and its internal state is approximately given by the thermal equilibrium state.
This local equilibrium ansatz is used to derive hydrodynamics in non-relativistic and relativistic fluids.
In Ref.\  \cite{hugo}, for example, relativistic ideal hydrodynamics with the abelian gauge interaction is derived 
by applying the variational principle and the Noether theorem to the Lagrangian which is represented by the ensemble of the fluid elements.

It is however known that the definition of a particle-like localized state is not straightforward 
in relativistic quantum systems \cite{newton-wigner,wightman1,wightman2}. 
Such a state is often represented by the non-local combination of Lorentz covariant fields, which is 
attributed to the difficulty to define localized state in the domain smaller than the Compton wave length.
See also recent papers \cite{odaka,kamefuchi1,kamefuchi2,pavsic,sa} and references therein.
In fact, the elementary degrees of freedom are known to be described by fields, not by particles in quantum field theory.
Because of this, the applicability of standard hydrodynamics 
should be considered carefully in the description of microscopic systems such as relativistic heavy-ion collisions.

One approach to avoid this difficulty is to consider coarse-graining based on the field degrees of freedom.
In the standard formulation of hydrodynamics, microscopic motions of constituent particles of a fluid are coarse-grained and used to define 
the internal energy of the fluid element. 
Applying thermodynamics to this internal energy, the macroscopic fluid motion is affected by coarse-grained degrees of freedom through pressure.
In the field theory, what is coarse-grained is the rapid oscillations of constituent fields of a fluid 
and thermodynamics is assumed to these oscillations.

The purpose of this paper is to study the formulation of a hydrodynamical model based on the field degrees of freedom.
For the sake of simplicity, we choose the 1+1 dimensional real scalar field as the constituent fields of a functional fluid.
We call this theory scalar ideal hydrodynamics.
Dynamics of the functional fluid is represented by the ensemble of the motions of the functional fluid elements defined in the 
function space of the scalar-field configuration.
The internal state of each element is assumed to be thermally equilibrated in the function space.
Thermodynamics in the function space which is obtained in Ref.\ \cite{scalar-SE} is utilized in the present work.
Our hydrodynamics is derived by the variation of an action with respect to the fluctuating trajectory of the functional fluid element.
This method is called the stochastic variational method (SVM) \cite{yasue, zambrini,koide_review_15,review_ucr,koide12,svm-field,koide_ucr1,koide_ucr2,koide_ucr3,kodama22,koide-curvedNSF}.
Scalar ideal hydrodynamics consists of the equations 
of the probability distribution of the field configuration (\ref{eqn:fih-1}), the internal energy (\ref{eqn:fih-2}) and the velocity functional (\ref{eqn:fih-3}). 
To find the correspondence relation to standard hydrodynamics, a further coarse-graining should be introduced. 
In this sense, our theory is regarded as a mesoscopic theory such as the Boltzmann equation 
in the dynamical hierarchy of many-body systems.
Scalar ideal hydrodynamics reproduces the exact behaviors of relativistic quantum field theory in a certain limit.
Indeed it is possible to define the deterministic trajectory associated with the motion of the fluid and then this trajectory 
is the functional generalization of the Bohmian trajectory. 
Therefore our theory may provide a new approach to study the influence of quantum-field theoretical fluctuation in collective flows 
of produced particles in relativistic heavy-ion collisions.

This paper is organized as follows.
The fluid element in the function space of the scalar-field configuration is introduced in Sec.\ \ref{sec:fluid element}.
The equation of the velocity functional is derived by applying the stochastic variational method in Sec.\ \ref{sec:sv}.
The energy and entropy conservation equations are phenomenological derived in Sec.\ \ref{sec:energy}.
The continuum limit of the derived equations is discussed in Sec.\ \ref{sec:continuum}.
In Sec.\ \ref{sec:mom_con}, the relation between the acceleration in the function space and the conserved momentum is studied.
In Sec.\ \ref{sec:nonlinearSEQ}, the functional Sch\"{o}dinger equation is shown to be reproduced from scalar ideal hydrodynamics in a certain limit.
Section \ref{sec:conc} is devoted to the summary and concluding remarks.

\section{Functional equilibrium ansatz} \label{sec:fluid element}

\begin{figure}[h]
\begin{center}
\includegraphics[scale=0.3]{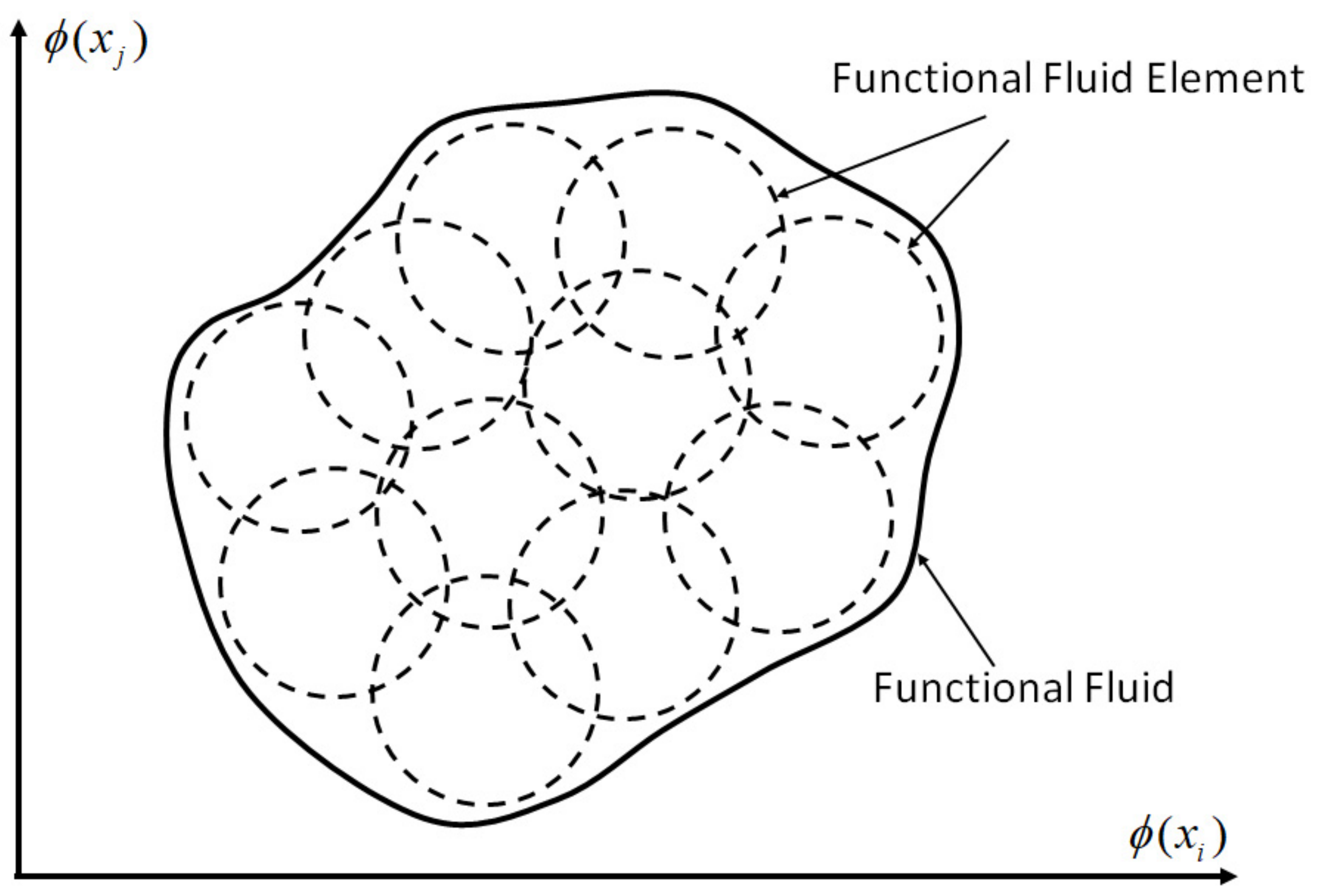}
\caption{The schematic figure of the distribution of the scalar-field configuration in the function space projected onto the two dimensional space spanned by $\phi(x_i)$ and $\phi(x_{j})$. 
The position $x_i$ is different from the position $x_j$.
The distribution of the functional fluid in this plane are represented by the ensemble of the functional fluid element denoted by the dotted circles.
}
\label{fig:functional_space}
\end{center}
\end{figure}

We consider the $1+1$ dimensional real scalar field $\phi(x,t)$.
The total spatial size (length) of the system is denoted by $l$, and the periodic boundary condition is applied, $\phi(x,t) = \phi(x+l,t)$.
The infinite limit of $l$ is taken at the last stage of calculations. 
This field $\phi(x,t)$ is still microscopic and any coarse-graining has not yet been introduced.

In principle, this field can oscillate in arbitrary scales but we are interested in coarse-grained behaviors where the rapid oscillations of $\phi(x,t)$ 
are already thermally equilibrated
\footnote{The scalar field can interact with other fields but such fields are assumed to be already thermally 
equilibrated, as is frequently assumed in the derivation of coarse-grained theories.}. 
To consider such a spatial coarse-graining, we introduce the minimum length scale which is denoted by $\Delta x$, and 
observe the behavior of the scalar field only at $2N$ discretized positions defined by 
\begin{eqnarray}
x_i = i (\Delta x) \,\,\,\,   (i = -N, -(N-1),\cdots , N-1) \, .
\end{eqnarray}  
The interval of the discretized positions is given by $\Delta x = l/(2N)$.
Therefore the field configuration is expressed as the ``position" in the $2N$-dimensional function space, 
\begin{eqnarray}
\underline{\phi} (t)= \left( \phi(x_{-N},t),\cdots,\phi(x_{N-1},t) \right)^{T} \, .
\end{eqnarray}
At a spatial position $x_i$, we cannot distinguish the fields in the domain $ x_i - \Delta x/2 \le x <x_i + \Delta /2$.
We suppose that the field in this domain is described statistically. 
For example, the distribution of the scalar field in this domain is given by $\rho_i (\phi (x_i))$.
Then the probability distribution of this coarse-grained field is 
\begin{eqnarray}
\rho_{cg} (\underline{\varphi}) = \int^\infty_{-\infty} [\ud \underline{\phi}] \prod^{N-1}_{i=-N} \rho_i (\phi (x_i))\delta (\varphi(x_i) - \phi (x_i)) \,  , \label{eqn:ini_dis}
\end{eqnarray}
where 
\begin{eqnarray}
[\ud \underline{\phi}] = \prod^{N-1}_{i=-N} \ud \phi (x_i) \, ,
\end{eqnarray}
and normalized by
\begin{eqnarray}
\int [\ud \underline{\varphi}] \rho_{cg} (\underline{\varphi}) = 1\, .
\end{eqnarray}
Because of this discretization, the spatial derivatives are identified with the differences, 
\begin{eqnarray}
\partial_x f(x_i) = \frac{f(x_{i+1})-f(x_{i-1})}{2\Delta x} \, ,
\end{eqnarray}
and 
\begin{eqnarray}
\partial^2_x f(x_i) = \frac{f(x_{i+2})+f(x_{i-2}) - 2 f(x_i)}{(2\Delta x)^2} \, ,
\end{eqnarray}
where $f(x)$ is an arbitrary function of $x$.

The fluid composed of the real scalar field distributes in a domain of the function space.
We then reexpress this distribution by the ensemble of functional fluid elements as is shown in Fig.\ \ref{fig:functional_space}. 
Suppose that each element has a small volume $\delta V_{\underline{\phi}}$ in the function space. 
The center of this volume is  
\begin{eqnarray}
\phi_{eff} (x_i) = \frac{1}{\delta V_{\underline{\phi}}} \int_{\delta V_{\underline{\phi}}} [\ud \underline{\phi}] \, \rho_{cg} (\underline{\phi}) \phi(x_i) \, .
\end{eqnarray}
Because $\underline{\phi}_{eff}(t)$ is regarded as the position of the functional fluid element in the $2N$-dimensional space, 
the motion of the functional fluid is represented by the ensemble of the trajectories of $\underline{\phi}_{eff}(t)$. 
Each fluid element has a specific mass which is a constant of motion in non-relativistic hydrodynamics,  
while it has a constant specific entropy in relativistic ideal hydrodynamics \cite{hugo}.
We apply the latter idea to derive functional ideal hydrodynamics.

The initial configuration of the functional fluid element is characterized by 
the initial probability distribution $\rho_0 (\underline{\phi_0})$ normalized by one,
\begin{eqnarray}
\int^\infty_{-\infty} [\ud \underline{\phi}_0] \, \rho_0 (\underline{\phi_0}) = 1 \, ,
\end{eqnarray}
where $\phi_0(x) = \phi_{eff}(x,t_I)$ represents the position of the functional fluid element at an initial time $t_I$. 
The probability distribution of the functional fluid element for arbitrary time $t>t_I$ is then defined by  
\begin{eqnarray}
\rho_F (\underline{\varphi}, t)  = \int^\infty_{-\infty} [\ud \underline{\phi}_0]  \, \rho_0 (\underline{\phi}_0)  \prod^{N-1}_{i=-N} \delta\left( \varphi (x_i ) - \phi_{eff}(x_i, t) \right)  \, . \label{eqn:rho_cla}
\end{eqnarray}
Thus the probability to find a functional fluid element in an infinitesimal volume $\delta V_{\underline{\varphi}}$ around $\underline{\varphi}$  
is given by $\rho_F (\underline{\varphi}, t) \delta V_{\underline{\varphi}}$.

To simplify complex dynamics associated with the coarse-grained degrees of freedom, 
we apply thermodynamics in the function space, which is developed in Ref.\ \cite{scalar-SE}.
Let us consider a static system consist of fields where all field configurations are distributed in a volume $V_f$ of the function space.
The energy and the entropy of this system are denoted by ${\cal E}$ and $S$, respectively.
From the first law of thermodynamics, ${\cal E}$ is expressed as a function of $V_f$ and $S$, 
\begin{eqnarray}
d{\cal E} = T dS - P_F dV_F \, . \label{eqn:ft}
\end{eqnarray}
Here we introduced the temperature $T$ and the pressure $P_F$ which is associated with the volume change in the function space. 
To derive functional hydrodynamics, we introduce the functional local equilibrium ansatz.
Let us introduce the internal energy distribution in the function space by 
\begin{eqnarray}
{\cal E} = \int_{V_F} [\ud \underline{\phi}] \, \varepsilon_{F}  \, .
\end{eqnarray}
Suppose that the effects of the rapid oscillations are assumed to be absorbed into the internal energy 
per unit functional fluid element, which is expressed by 
\begin{eqnarray}
\frac{\varepsilon_{F} (\rho_F, S_F)}{\rho_F} \, ,
\end{eqnarray}
where $S_F$ denotes the specific entropy of the functional fluid element. See also the discussion in Refs.\ \cite{hugo,scalar-SE}.
The functional local equilibrium ansatz means that the change of this energy is determined by the thermodynamical relation obtained from Eq.\ (\ref{eqn:ft}), 
\begin{eqnarray}
\ud \frac{\varepsilon_F}{\rho_F} = T \ud S_F - P_F \ud \rho^{-1}_F \, . \label{eqn:functional-thermodinamics}
\end{eqnarray}
This relation is used to derive the equation of the entropy distribution later.

In the variational derivation of non-relativistic ideal hydrodynamics, 
the Lagrangian is given by the contributions of 
the kinetic and internal energies of the fluid elements defined in the Lagrange coordinates \cite{review_ucr,koide12}.
Therefore the motion of the fluid element becomes free-streaming when the internal energy term is ignored.
A similar structure is assumed in the function space: the motion of the functional fluid element is given 
by the free scalar-field equation in the case of no internal energy.
The Lagrangian satisfying this condition is defined by  
\begin{eqnarray}
L 
&=&  \int^\infty_{-\infty} [\ud \underline{\phi}_0] \, 
\rho_0 (\underline{\phi}_0) \left[  (\Delta x) \sum_{i=-N}^{N-1} \left\{ \frac{1}{2c^2} (\partial_t \phi_{eff} (x_i,t))^2 - \frac{1}{2} (\partial_x \phi_{eff}(x_i,t))^2 - V(\phi_{eff}(x_i,t))  \right\} \right. \nonumber \\
&&-  \left. \frac{\varepsilon_{F}(\rho_F (\underline{\phi}_{eff}(t),t),S_F) }{\rho_F (\underline{\phi}_{eff}(t),t)} \right]
\, , \label{eqn:cla-lag}
\end{eqnarray}
where $V(\phi_{eff}(x_i,t))$ denotes the mass term.
We can however introduce non-linear field interactions in $V(\phi_{eff}(x_i,t))$ to consider solitons \cite{lee,nugaev,glendenning}.

Scalar ideal hydrodynamics is derived by applying a variational method to the action defined by this Lagrangian.
When hydrodynamics is applied to microscopic systems, we should utilize the stochastic variational method.

\section{Stochastic variation} \label{sec:sv}

In the standard variational method,
it is implicitly assumed that virtual trajectories are always differentiable, but 
its applicability is not obvious. 
In fact, there exists a generalized framework of variation, the stochastic variational method (SVM), 
where non-differentiable virtual trajectories are considered in the optimization processes \cite{yasue, zambrini,koide_review_15,review_ucr,koide12,svm-field,koide_ucr1,koide_ucr2,koide_ucr3,kodama22,koide-curvedNSF}.
As review papers of SVM, see, for example, Refs.\ \cite{zambrini,koide_review_15,review_ucr,kodama22}. 
Reference\ \cite{kuipers2} is also useful to see the related background. 
SVM is the natural generalization of the classical variational method and thus the optimized result coincides with 
that of the classical variation in the smoothing limit of the virtual trajectories.
In this approach, non-differentiability is considered to be the indispensable property of microscopic systems. 
For example, the Schr\"{o}dinger equation is derived by applying the stochastic variation to the action 
which leads to the Newton equation under the application of the classical variation \cite{yasue}.
Moreover, viscous hydrodynamics is obtained by applying SVM to the action 
which gives ideal hydrodynamics under the classical variation \cite{koide12,review_ucr}.
Therefore SVM should be applied to derive coarse-grained dynamics in microscopic systems.

There are various formulations of SVM. 
In this paper, we apply the theory developed by Yasue \cite{yasue}.
As other formulations of SVM, see Refs.\ \cite{zambrini,review_ucr,kuipers} and references therein.
So far, SVM has been applied exclusively to particle systems and continuum medium (See, however, Ref.\ \cite{svm-field}). 
Here we generalize this idea to a field-theoretical system.
Then the variation is applied to the trajectory of the functional fluid element which fluctuates because of the interaction with other elements, its internal excitation and so on.

The typical example of the non-differentiable trajectory is known in Brownian motion.
In SVM, we thus assume that the trajectory of the functional fluid element is characterized by the forward ($\ud t >0$) and backward ($\ud t <0$) 
stochastic differential equations (SDE's) \cite{svm-field}, 
\begin{eqnarray}
\ud \widehat{\phi}_{eff} (x_i,t) &=& u_+ (x_i,  \underline{\widehat{\phi}}_{eff} ,t ) \ud t 
+ \sqrt{\frac{2\nu}{\Delta x}} \ud \widehat{W}_+ (x_i,t) \,\,\,\,\, (\ud t > 0 ) \, , \label{eqn:fsde}\\
\ud \widehat{\phi}_{eff} (x_i, t) &=& u_-  ( x_i, \underline{\widehat{\phi}}_{eff},t ) \ud t 
+ \sqrt{\frac{2\nu}{\Delta x}} \ud \widehat{W}_- (x_i,t) \,\,\,\,\, (\ud t < 0 ) \, . \label{eqn:bsde}
\end{eqnarray}
Here we used $(\,\widehat{\,\,}\,)$ to represent stochastic quantities and $\ud \widehat{A}(x,t)= A(x,t+\ud t) - A(x,t)$ for a stochastic field $\widehat{A}(x,t)$. 
The standard Wiener processes $\widehat{W}_\pm (x_i,t)$ satisfy the following correlations: 
\begin{eqnarray}
\uE \left[ \ud \widehat{W}_\pm (x_i,t)\right] &=& 0 \, ,\\
\uE \left[ \ud \widehat{W}_\pm (x_i,t) \ud \widehat{W}_\pm (x_j,t^\prime) \right] &=& |\ud t| \delta_{i,j} \delta_{t,t^\prime}  \, ,
\end{eqnarray}
where $\uE [\,\,\,]$ denotes the ensemble average for the Wiener processes \cite{book_gardiner}. 
The velocity functionals $u_\pm (x_i, \underline{\varphi},t)$ are assumed to be smooth functionals of $\underline{\varphi}$, 
but $u_\pm ( x_i, \underline{\widehat{\phi}}_{eff},t )$ are stochastic because of $\underline{\widehat{\phi}}_{eff}$.
The unknown functionals $u_\pm (x_i,\underline{\phi},t)$ are determined by the stochastic optimization.

The introduction of the two SDE's is attributed to the non-differentiability of stochastic trajectories. 
Here the standard definition of velocity is not applicable because the left and right-hand limits of the inclination of stochastic trajectories do not agree.
Because of this ambiguity of velocity, Nelson introduced two time derivatives \cite{nelson}: one is the mean forward derivative $\uD_+$ and the other the mean backward derivative $\uD_-$, which are, respectively, defined by 
\begin{eqnarray}
\uD_+ \widehat{\phi}_{eff} (x_i, t) 
&=& \lim_{\ud t \rightarrow 0+} \uE \left[ \frac{\widehat{\phi}_{eff} (x_i, t+\ud t)- \widehat{\phi}_{eff} (x_i, t)}{\ud t}
\Big| {\cal P}_t \right] \, ,\\
\uD_- \widehat{\phi}_{eff} (x_i, t) 
&=& \lim_{\ud t \rightarrow 0-} \uE \left[ \frac{\widehat{\phi}_{eff} (x_i, t+\ud t)- \widehat{\phi}_{eff} (x_i, t)}{\ud t}
\Big| {\cal F}_t \right] \, .
\end{eqnarray}
These expectation values are conditional averages where ${\cal P}_t$ (${\cal F}_t$) indicates to fix values of $\widehat{\phi}_{eff} (x_i,t^\prime)$ 
for $t^\prime \le t$ ($t^\prime \ge t$). 
For the $\sigma$-algebra of all measurable events of $\widehat{\phi}_{eff} (x_i,t)$,  ${\cal P}_t$ and ${\cal F}_t$ represent an increasing and a decreasing family of sub-$\sigma$-algebras, respectively.
Applying these definitions to Eqs.\ (\ref{eqn:fsde}) and (\ref{eqn:bsde}), 
we find 
\begin{eqnarray}
\uD_\pm \underline{\widehat{\phi}}_{eff} (t)  &=& \underline{u}_\pm ( \underline{\widehat{\phi}}_{eff}(t) ,t ) \, , 
\end{eqnarray}
where
\begin{eqnarray}
\underline{u}_\pm (\underline{\phi}_{eff},t) = \left( u_\pm (x_{-N},\underline{\phi}_{eff},t ), \cdots, u_\pm (x_{N-1},\underline{\phi}_{eff},t ) \right)^T \, .
\end{eqnarray}

The stochastic motion of the functional fluid element should satisfy both of the forward and backward SDE's and thus 
these unknown functionals $\underline{u}_\pm (\underline{\phi}_{eff}, t)$ are not independent. 
To see this, we introduce the probability distribution of the functional fluid element by 
\begin{eqnarray}
\rho_F (\underline{\varphi}, t)  = \int^\infty_{-\infty} [\ud \underline{\phi}_0]  \, \rho_0 (\underline{\phi}_0)  \prod^{N-1}_{i=-N} 
\uE \left[ \delta\left( \varphi (x_i ) - \widehat{\phi}_{eff}(x_i, t) \right) \right] \, . \label{eqn:rho_sto}
\end{eqnarray}
One can easily see that this is the stochastic generalization of Eq.\ (\ref{eqn:rho_cla}).
When we adapt the forward SDE to the above definition, the functional Fokker-Planck equation is given by 
\begin{eqnarray}
\partial_t \rho_F ( \underline{\varphi} ,t ) 
=
- \partial_{\underline{\varphi}} *  \{\underline{u}_+ ( \underline{\varphi},t )\rho_F ( \underline{\varphi}, t )\}
+ \frac{\nu}{\Delta x}
\partial^2_{\underline{\varphi}} 
\rho_F ( \underline{\varphi}, t ) \, , \label{eqn:ffp}
\end{eqnarray}
where we introduced the following notations: 
\begin{eqnarray}
\underline{A} * \underline{B} &=& \sum_{i=-N}^{N-1} A(x_i) B(x_i) \, , \\
\partial_{\underline{\varphi}} 
&=& \left(  \frac{\partial}{\partial \varphi(x_{-N})} , \cdots, \frac{\partial}{\partial \varphi(x_{N-1})}  \right)^T \, , \\
\partial^2_{\underline{\varphi}} 
&=& \partial_{\underline{\varphi}}  * \partial_{\underline{\varphi}} \, .
\end{eqnarray}
At the same time, when the backward SDE is applied, another equation is obtained, 
\begin{eqnarray}
\partial_t \rho_F ( \underline{\varphi} ,t ) 
=
- \partial_{\underline{\varphi}} * \{ \underline{u}_- ( \underline{\varphi},t )\rho_F ( \underline{\varphi}, t )\}
- \frac{\nu}{\Delta x}
\partial_{\underline{\varphi}}^2 
\rho_F ( \underline{\varphi}, t ) \, .\label{eqn:bfp}
\end{eqnarray}
These two equations should be equivalent and then we find that the unknown smooth functionals 
should satisfy the functional consistency condition \cite{svm-field},
\begin{eqnarray}
u_+ (x_i, \underline{\varphi}, t) - u_- (x_i, \underline{\varphi}, t)
= \frac{2\nu}{\Delta x} \frac{\partial}{\partial \varphi(x_i)} \ln \rho_F ( \underline{\varphi}, t ) 
\, . \label{eqn:cc}
\end{eqnarray}
Using this condition, 
we can show that the above two Fokker-Planck equations are reduced to the same equation of continuity in the function space,
\begin{eqnarray}
\partial_t \rho_F ( \underline{\varphi} ,t ) 
= - \partial_{\underline{\varphi}} * \{\rho_F ( \underline{\varphi}, t )\, \underline{v} ( \underline{\varphi},t )\} \, , \label{eqn:feoc}
\end{eqnarray}
where the mean velocity functional is introduced by 
\begin{eqnarray}
 \underline{v} ( \underline{\varphi},t )
=
\left( v(x_{-N}, \underline{\varphi},t), \cdots, v(x_{N-1}, \underline{\varphi},t) \right)^T
= \frac{ \underline{u}_+ ( \underline{\varphi},t ) +  \underline{u}_- ( \underline{\varphi},t )}{2} \, .
\label{eqn:def_v}
\end{eqnarray}

To apply SVM to derive scalar ideal hydrodynamics, 
$\phi_{eff}(x,t)$ in the classical Lagrangian (\ref{eqn:cla-lag}) should be replaced with the corresponding stochastic quantities, $\widehat{\phi}_{eff}(x,t)$.
Because of the two time derivatives $\uD_\pm$ introduced above, this replacement is ambiguous.
In this work, the time derivative term in the classical Lagrangian 
is replaced by the average of the two contributions,  $\uD_\pm$, leading to 
\begin{eqnarray}
L_{sto}
&=&  
 \int^\infty_{-\infty} [\ud \underline{\phi}_0] \, 
\rho_0 (\underline{\phi}_0) \left[  (\Delta x) \sum_{i=-N}^{N-1} \left\{ \frac{1}{2c^2} \frac{(\uD_+\widehat{\phi}_{eff} (x_i,t))^2 + (\uD_-\widehat{\phi}_{eff} (x_i,t))^2 }{2} \right. \right. \nonumber\\
&& \left. \left.- \frac{1}{2} (\partial_x \widehat{\phi}_{eff}(x_i,t))^2 - V(\widehat{\phi}_{eff}(x_i,t))  \right\} 
- \frac{\varepsilon_{F}(\rho_F (\widehat{\underline{\phi}}_{eff}(t),t),S_F) }{\rho_F (\widehat{\underline{\phi}}_{eff}(t),t)} \right]
\, . \label{eqn:sto-lag}
\end{eqnarray}
The effect of viscosity is introduced by considering more general quadratic forms in the replacement of the time derivative terms, 
but this is not considered in the present work \cite{koide12,review_ucr,koide-curvedNSF,koide_ucr2,koide_ucr1}.
The specific entropy $S_F$ behaves as a constant of motion in the functional Lagrange coordinates and thus its variation is not considered.

What we can optimize is only the mean behavior of stochastic systems. 
Therefore the action is defined by the expectation value of the stochastic Lagrangian, 
\begin{eqnarray}
I = \int^{t_F}_{t_I} \ud t \, \uE \left[ L_{sto} \right] \, ,
\label{eqn:sto_act}
\end{eqnarray}
where $t_F (> t_I)$ is a final time.
The variation of the functional fluid element is defined by 
\begin{eqnarray}
\widehat{\phi}_{eff} (x_i ,t) \longrightarrow \widehat{\phi}_{eff} (x_i ,t) + \delta f(x_i, \widehat{\underline{\phi}}_{eff}(t),t) \, .
\label{eqn:variation_phi}
\end{eqnarray}
Here $\delta f(x_i, \underline{\varphi},t)$ is an arbitrary smooth functional satisfying 
\begin{eqnarray}
 \delta f(x_i, \underline{\varphi},t_I) =  \delta f(x_i, \underline{\varphi},t_F ) = 0 \, ,
\end{eqnarray}
and   
\begin{eqnarray}
 \delta f(x_{-N},\underline{\varphi},t) =  \delta f(x_{N},\underline{\varphi},t )  \, .
\end{eqnarray}
The stochastic variation of the action then leads to
\begin{eqnarray}
\delta I_{sto} 
&=& 
-\int^{t_F}_{t_I} \ud t  \, \sum_{i=-N}^{N-1}
\uE \left[
 \uD_- \frac{\partial L_{sto}}{\partial (\uD_+ \widehat{\phi}_{eff}(x_i, t))}
+
 \uD_+ \frac{\partial L_{sto}}{\partial (\uD_- \widehat{\phi}_{eff}(x_i,t))}
+
 \partial_x \frac{\partial L_{sto}}{\partial (\partial_x \widehat{\phi}_{eff}(x_i,t))} \right. \nonumber \\
&& \left. - \frac{\partial L_{sto}}{\partial \widehat{\phi}_{eff}(x_i,t)}
- \frac{1}{\rho_F}  \frac{\partial}{\partial \widehat{\phi}_{eff}(x_i,t)} \frac{\partial  L_{sto}}{\partial \rho^{-1}_F}
\right]  \delta f(x_i, \underline{\phi}_{eff}(t),t) \, .
\end{eqnarray}
In the last term, we used 
\begin{eqnarray}
\lefteqn{\delta \int^\infty_{-\infty} [\ud \underline{\phi}_0] \rho_0 (\underline{\phi}_0) \frac{\varepsilon_f (\rho_F (\widehat{\underline{\phi}}_{eff}(t),t), S_F)}{\rho_F (\widehat{\underline{\phi}}_{eff},t)} } && \nonumber \\
&=& \left. -\int^\infty_{-\infty} [\ud \underline{\phi}_0]  \frac{ \rho_0 (\underline{\phi}_0)}{\rho_F (\underline{\varphi},t)}   \delta \underline{f} (\underline{\varphi},t) * \partial_{\underline{\varphi}}  
\left( \frac{\partial}{\partial \rho^{-1}_F} \frac{\varepsilon_f (\rho_F (\underline{\varphi},t),S_f)}{\rho_F (\underline{\varphi},t)} \right)_{S_F}
\right|_{\underline{\varphi}= \widehat{\underline{\phi}}_{eff}(t)} \, .
\label{eqn:vari_int_ener}
\end{eqnarray}
See the discussions in Refs.\ \cite{koide12,review_ucr} and Appendix \ref{app:pressure} for more detail.
In SVM, we require that $\delta I_{sto}$ disappears for an arbitrary choice of $\delta f (x_i, \underline{\varphi},t)$ and 
an arbitrary value of $\widehat{\underline{\phi}}_{eff}(t)$.
To satisfy these conditions, the unknown smooth functionals $\underline{u}_\pm (\underline{\varphi},t)$ are given by the solutions of the following equation:
\begin{eqnarray}
&& \left[
\frac{ \uD_- u_+ (x_i,\widehat{\underline{\phi}}_{eff}(t),t) 
+  \uD_+ u_- (x_i,\widehat{\underline{\phi}}_{eff}(t),t)}{2c^2} 
- \partial^2_{x} \widehat{\phi}_{eff}(x_i,t)  \right. \nonumber \\
&&  \left. + \frac{\partial V (\widehat{\phi}_{eff}(x_i,t)) }{\partial \widehat{\phi}_{eff}(x_i,t)} 
  - \frac{1}{\rho_f} \frac{\partial}{\partial \widehat{\phi}_{eff}(x_i,t)} 
\left( \frac{\partial}{\partial \rho^{-1}_F} \frac{\varepsilon_F (\rho_f, S_f)}{\rho_F} \right)_{S_F}
\right]_{\widehat{\underline{\phi}}_{eff}(t) = \underline{\varphi}} = 0 \, . 
\end{eqnarray}
Here $|_{\widehat{\underline{\phi}}_{eff}(t) = \underline{\varphi}}$ indicates that the stochastic quantity $\widehat{\underline{\phi}}_{eff}(t) $ is replaced with the variable $\underline{\varphi}$ in the last stage of calculations.
The acceleration terms are calculated by using Ito's lemma \cite{book_gardiner},
\begin{eqnarray}
\uD_\pm u_\mp (x_i, \widehat{\phi}_{eff} (t),t) = 
\left.
\left\{ \partial_t 
+ \underline{u}_\pm (\underline{\varphi},t)* \partial_{\underline{\varphi}}   
\pm \frac{\nu}{\Delta x} \partial^2_{\underline{\varphi}}\right\}  u_\mp (x_i, \varphi(t),t)
\right|_{\underline{\varphi}= \widehat{\underline{\phi}}_{eff}(t)} \, .
\end{eqnarray}
In the end, the result of the stochastic variation is represented by
\begin{eqnarray}
&& \partial_t  v(x_i, \underline{\varphi}, t) + \underline{v}(\underline{\varphi}, t) * \partial_{\underline{\varphi}}  v (x_i, \underline{\varphi}, t)
- \frac{2\nu^2}{ (\Delta x)^2}   \frac{\partial}{\partial \varphi(x_i)} \left( \frac{1}{\sqrt{\rho_F (\underline{\varphi},t)}} \partial_{\underline{\varphi}}^2 \sqrt{\rho_F (\underline{\varphi},t)} \right) 
\nonumber \\
&& 
- c^2 \partial^2_x \varphi (x_i) 
+ c^2 \frac{\partial  V(\varphi(x_j))}{\partial \varphi (x_i)}
+ c^2 \frac{1}{\Delta x}\frac{1}{\rho_F (\underline{\varphi},t)} \frac{\partial P_{F} (\rho_F, S_F)}{\partial \varphi(x_i)} 
 = 0  \, ,
\label{eqn:fluid_qft_pressure}
\end{eqnarray}
where 
\begin{eqnarray}
P_F (\rho_F,S_F)= - \left( \frac{\partial}{\partial \rho^{-1}_F} \frac{\varepsilon_F (\rho_F (\underline{\varphi},t), S_F)}{\rho_F (\underline{\varphi},t)} \right)_{S_F} \, .
\label{eqn:pressure}
\end{eqnarray}
This quantity, $P_F (\rho_F,S_F)$, is shown to be interpreted as the pressure in the function space using 
the first law of thermodynamics (\ref{eqn:functional-thermodinamics}).
The structure of Eq.\ (\ref{eqn:fluid_qft_pressure}) can be regarded as the functional generalization of the momentum equation of non-relativistic ideal hydrodynamics.

The last term on the first line of Eq.\ (\ref{eqn:fluid_qft_pressure}) represents the correction term to the functional pressure induced 
by the fluctuation of the functional fluid element since this term is proportional to the square of the noise intensity $\nu^2$.  
In fact, Eq.\ (\ref{eqn:fluid_qft_pressure}) can be reexpressed as 
\begin{eqnarray}
&& \partial_t  v(x_i, \underline{\varphi}, t) + \underline{v}(\underline{\varphi}, t) * \partial_{\underline{\varphi}}  v (x_i, \underline{\varphi}, t)
- c^2 \partial^2_x \varphi (x_i) 
+ c^2 \frac{\partial  V(\varphi(x_j))}{\partial \varphi (x_i)} \nonumber \\
&& + c^2 \frac{1}{\Delta x}\frac{1}{\rho_F (\underline{\varphi},t)}\sum_{j=1}^{N-1} \frac{\partial P^{ij}_{F} (\rho_F, S_F)}{\partial \varphi(x_j)} 
 = 0  \, , \label{eqn:fluid_qft_pressure2}
\end{eqnarray}
where the functional stress tensor is defined by 
\begin{eqnarray}
P^{ij}_F (\rho_F, S_F) = P_F (\rho_F, S_F) \delta^{ij}
- \frac{\nu^2}{ c^2 (\Delta x)}  \rho_f (\underline{\varphi},t)  \frac{\partial}{\partial \varphi(x_i)} \frac{\partial}{\partial \varphi(x_j)} \ln \rho_F (\underline{\varphi},t) \, . 
\label{eqn:eff-pressure}
\end{eqnarray}
We see later that this term reproduces the exact fluctuation of the quantized scalar field when the noise intensity $\nu$ is chosen appropriately.

The structure of Eq.\ (\ref{eqn:fluid_qft_pressure}) is found even by considering the moment equations in the functional phase space.
See the discussion in Appendix \ref{app:kinetic}.

\subsection{classical limit and functional Bohmian trajectory}
\label{sec:bohm}

So far, we have considered the stochastic motion of the functional fluid element. 
We introduce however the deterministic scalar field, $\phi_{det}(x,t)$, which is defined by solving 
\begin{eqnarray}
\partial_t  \phi_{det}(x_i,t) =  v(x_i, \underline{\phi}_{det}(t), t) \, . \label{eqn:tra}
\end{eqnarray}
Substituting $\phi_{det}(x,t)$ into Eq.\ (\ref{eqn:fluid_qft_pressure2}), the evolution equation of $\phi_{det}(x,t)$ is given by 
\begin{eqnarray}
&& \partial^2_t   \phi_{det}(x_i,t)
- c^2 \partial^2_x  \phi_{det}(x_i,t)
+ c^2 \frac{\partial  V( \phi_{det}(x_j,t))}{\partial  \phi_{det}(x_i,t)} \nonumber \\
&& + c^2 \frac{1}{\Delta x}\frac{1}{\rho_F (\underline{\phi}_{det}(t),t)}\sum_{j=1}^{N-1} \frac{\partial P^{ij}_{F} (\rho_F, S_F)}{\partial  \phi_{det}(x_i,t)} 
 = 0  \, , 
\label{eqn:bohm_hydro}
\end{eqnarray}
where we used
\begin{eqnarray}
\partial^2_t   \varphi_{det}(x_i,t) = \frac{\ud}{\ud t} v(x_i, \underline{\varphi}_{det}(t), t) 
=\left.  \left\{ \partial_t + \underline{v} (\underline{\varphi}, t) * \partial_{\underline{\varphi}} \right\} v(x_i, \underline{\varphi}, t)
\right|_{\underline{\varphi} = \underline{\phi}_{det}(t)} \, .
\end{eqnarray}
The first three terms on the left-hand side represent the classical equation of the scalar field.
The last term involves the functional gradient of $P_F$ and the term proportional to $\nu^2$. The the latter contribution does not appear 
in the classical variation.
That is, 
the SVM optimization reproduces the result of the classical variation
in the vanishing limit of the fluctuation of the functional fluid element.

The trajectory of $\phi_{det}(x_i,t)$ is reminiscent of the Bohmian trajectory.
It is sometimes considered that the results of quantum mechanics can be represented by the ensemble of particles which have precise positions. 
The paths of these particles are called the Bohmian trajectories, which are determined by solving {\it the Newton equation with the quantum potential} \cite{book_holland}.
As shown later, Eq.\ (\ref{eqn:fluid_qft_pressure2}) (that is, Eq.\ (\ref{eqn:bohm_hydro})) reproduces the exact behaviors of relativistic quantum field theory 
when the contribution from the functional pressure is not considered. 
That is, the path of $\phi_{det}(x,t)$ is described by {\it the classical field equation with the functional quantum potential} 
and, in this sense, is regarded as the functional generalization of the Bohmian trajectory.
This path however does not describe the motion of a quantized particle and thus is not the complete analogue of the Bohmian 
trajectory in quantum mechanics. 
The Bohmian trajectory based on the motion of a quantized particle is studied in Ref.\ \cite{durr}.

\section{Conservation laws of energy and entropy } \label{sec:energy}

From the stochastic Lagrangian (\ref{eqn:sto-lag}), the total energy of this system is given by 
\begin{eqnarray}
{\cal E}
&=&  \int^\infty_{-\infty} [\ud \underline{\phi}_0] \, 
\rho_0 (\underline{\phi}_0) \uE \left[  (\Delta x) \sum_{i=0}^{N-1} \left\{ \frac{1}{2c^2} \frac{(\uD_+\widehat{\phi}_{eff} (x_i,t))^2 + (\uD_-\widehat{\phi}_{eff} (x_i,t))^2 }{2} \right. \right. \nonumber\\
&& \left. \left.+ \frac{1}{2} (\partial_x \widehat{\phi}_{eff}(x_i,t))^2 + V(\widehat{\phi}_{eff}(x_i,t))  \right\} 
+ \frac{\varepsilon_{F}(\rho_F(\widehat{\underline{\phi}}_{eff}(t),t),S_F) }{\rho_F (\widehat{\underline{\phi}}_{eff}(t),t)} \right] \, .
\end{eqnarray}
The time evolution of $\varepsilon_F$ is determined so that ${\cal E}$ is conserved.
Using Eqs.\ (\ref{eqn:feoc}) and (\ref{eqn:fluid_qft_pressure}),
this time derivative is shown to be given by 
\begin{eqnarray}
\frac{\ud {\cal E}}{\ud t}
&=& 
 \int [\ud \underline{\varphi}]  \left\{ P_{F} (\rho_F, \varepsilon_F)
  \partial_{\underline{\varphi}} * \underline{v}(\underline{\varphi},t) 
+ \partial_t\varepsilon_{F} (\underline{\varphi},t)  \right\} 
\, .
\end{eqnarray}
Here $\varepsilon_F$ is expressed as a functional of $\underline{\varphi}$.
To satisfy the energy conservation, the right-hand side should be expressed by the integral of the divergence 
of the energy current $\underline{J}$ in the function space.
When the energy current $\underline{J}$ is assumed to be given by $\varepsilon_F (\underline{\varphi},t) \underline{v}(\underline{\varphi},t)$,  
the evolution equation of the internal energy distribution is given by  
\begin{eqnarray}
 \partial_t\varepsilon_{F} (\underline{\varphi},t) 
&=& - \partial_{\underline{\varphi}} * ( \varepsilon_F (\underline{\varphi},t)  \underline{v}(\underline{\varphi},t)  ) -   P_{F}(\rho_F, \varepsilon_F)
  \partial_{\underline{\varphi}} * \underline{v}(\underline{\varphi},t)  \, .
  \label{eqn:ener}
\end{eqnarray}
It is easy to see that this equation is the functional generalization of the energy equation of non-relativistic ideal hydrodynamics 
\cite{book_landau_lifshitz}.

Using this, we can show the conservation of the entropy \cite{scalar-SE}.
The Euler equation of thermodynamics in the function space is given by 
\begin{eqnarray}
\varepsilon_F = T s_F - P_F \, , \label{eqn:ther_euler}
\end{eqnarray}
where $s_F$ denotes the entropy distribution in the function space.
Then, from the first law (\ref{eqn:ft}), the change of the entropy distribution on a functional fluid element should be described by  
\begin{eqnarray}
\frac{\ud s_F}{\ud t_{\varphi}}  = \frac{1}{T} \frac{\ud \varepsilon_F}{\ud t_{\varphi}}  \, ,
\end{eqnarray}
where the material derivative in the function space is introduced, 
\begin{eqnarray}
\frac{\ud}{\ud t_{\varphi}}  = \partial_t + \underline{v} * \partial_{\underline{\varphi}} \, .
\end{eqnarray} 
Substituting Eq.\ (\ref{eqn:ener}) into the right-hand side, we can show that the entropy distribution satisfies the equation of continuity in the function space,
\begin{eqnarray}
\partial_t s_F (\underline{\varphi},t) = - \partial_{\underline{\varphi}} * (s_F (\underline{\varphi},t) \underline{v}(\underline{\varphi},t) ) \, .
\end{eqnarray}
See also the discussion in Ref.\ \cite{scalar-SE}.

\section{Continuum limit} \label{sec:continuum}

Let us consider the continuum limit where $\Delta x \longrightarrow 0$ and $l \longrightarrow \infty$.
The spatial integral and the functional derivative are then introduced by 
\begin{eqnarray}
\Delta x \sum_{i=-N}^{N-1} &\longrightarrow& \int^\infty_{-\infty} \ud x \, , \\
\frac{1}{\Delta x} \frac{\partial}{\partial \varphi (x_i)} &\longrightarrow& \frac{\delta}{\delta \varphi(x)} \, .
\end{eqnarray}
Suppose that the functional hydrodynamical quantities are well defined in this limit, 
\begin{eqnarray}
\rho_F (\underline{\varphi},t) &\longrightarrow& \rho_F [\varphi,t] \, , \\
\varepsilon_F  (\underline{\varphi},t) &\longrightarrow& \varepsilon_F [\varphi,t] \, , \\
v(x_i ,\underline{\varphi}, t) &\longrightarrow&  v [x, \varphi, t] \, .
\end{eqnarray}
Under this assumption, scalar ideal hydrodynamics is summarized as 
\begin{eqnarray}
\partial_t \rho_F [ \varphi ,t ] 
&=& - \int^\infty_{-\infty} \ud x\, \frac{\delta}{\delta \varphi(x)}   \left(  \rho_F [ \varphi, t ] v [ x, \varphi,t ]  \right) \, ,\label{eqn:fih-1} \\
 \partial_t\varepsilon_{F} [ \varphi ,t ] 
&=& -  \int^\infty_{-\infty} \ud x\, \frac{\delta}{\delta \varphi(x)}\left(  \varepsilon_F [\varphi,t] v [x, \varphi,t] \right)
-  \int^\infty_{-\infty} \ud x\,   
P_F (\varepsilon_F, \rho_F) \frac{\delta}{\delta \varphi(x)}  v [x, \varphi,t] 
\, ,\label{eqn:fih-2} \\
\partial_t  v [x, \varphi, t] 
&=& - \int^\infty_{-\infty} \ud y\, v [y, \varphi, t ] \frac{\delta}{\delta \varphi(y)}   v [ x, \varphi, t ]
+ c^2 \partial^2_x \varphi (x) \nonumber \\
&& - c^2 \frac{\delta}{\delta \varphi (x)} \int^\infty_{-\infty} \ud y\, V(\varphi(y))
- c^2 \frac{1}{\rho_F [\varphi,t]} \frac{\delta P_F (\varepsilon_F, \rho_F)}{\delta \varphi(x)}
\nonumber \\
&&
+ 2\nu^2 \int^\infty_{-\infty} \ud y\,    \frac{\delta}{\delta \varphi(x)} \left( \frac{1}{\sqrt{\rho_F [\varphi,t]}}  \frac{\delta^2}{\delta^2 \varphi(y)} \sqrt{\rho_F [\varphi,t]} \right) \, .
\label{eqn:fih-3}
\end{eqnarray}
It should be emphasized that the existence of this limit is not trivial.
For example, the second-order functional derivative $\delta^2/\delta^2 \varphi(y)$ is known to cause a singular behavior. 
This singularity appears even in quantum field theory but does not cause any problem when observables are calculated by, for example, introducing the lattice discretization \cite{sym,lus}.

\section{Momentum conservation} \label{sec:mom_con}

The first two equations (\ref{eqn:fih-1}) and (\ref{eqn:fih-2}) in scalar ideal hydrodynamics 
correspond to the conservation laws of the probability and the energy in the function space.
The last equation is related to the momentum conservation but not equivalent. 
It is because $v(x_i, \underline{\varphi}, t)$ characterizes the motion 
in the function space and is not the fluid velocity in the spatial configuration space.

To find the conserved momentum in this system, we apply the Noether theorem associated with the spatial translation to the stochastic action \cite{review_ucr,koide_ucr1,misawa},
\begin{eqnarray}
x_i \longrightarrow x^\prime_i  = x_i + \epsilon \, ,
\end{eqnarray}
where $\epsilon$ is an infinitesimal constant.
Requiring that the stochastic action (\ref{eqn:sto_act}) is invariant for this translation, 
the conserved momentum is defined by the expectation value of the momentum functional,
\begin{eqnarray}
\frac{\ud}{\ud t }  \sum_{i=-N}^{N-1} \int^\infty_{-\infty} [\ud \underline{\varphi}] \rho_F (\underline{\varphi},t) p(x_i, \underline{\varphi},t) = 0 \, .
\end{eqnarray}
where the momentum functional is defined by 
\begin{eqnarray}
p (x_i, \underline{\varphi},t) 
= - \frac{1}{c^2} v ( x_i, \underline{\varphi},t ) \partial_x \varphi (x_i)
\, .  \label{eqn:mom_func}
\end{eqnarray}
One can see that the conserved momentum is related to $v ( x_i, \underline{\varphi},t )$, although it is not the velocity in the spatial configuration space.

To find the momentum equation in the spatial configuration space, 
we introduce the momentum {\it function} which is obtained by the expectation value of the momentum functional,  
\begin{eqnarray}
\langle p(x_i,t) \rangle_F = \int^\infty_{-\infty} [\ud \underline{\varphi}] \, \rho_F (\underline{\varphi},t)  \,
p (x_i, \underline{\varphi},t) \, .
\end{eqnarray}
This quantity satisfies the following differential equation: 
\begin{eqnarray}
\partial_t \langle p (x_i, t) \rangle_F 
&=& 
- \partial_x  \left\{  \frac{1}{2c^2}\langle v^2 (x_i,t) \rangle_F 
+
\int^\infty_{-\infty} [\ud \underline{\varphi}] \, \rho_F (\underline{\varphi},t) \left(
\frac{1}{2} (\partial_x \varphi(x_i) )^2 
- V(\varphi(x_i))
\right)
\right\} \nonumber \\
&& 
-  c^{-2}  \frac{\nu^2}{ (\Delta x)^2} \int^\infty_{-\infty}  [\ud \underline{\varphi}]    (\partial_x \varphi (x_i))  
\partial_{\underline{\varphi}} * \rho_F (\underline{\varphi},t) 
\partial_{\underline{\varphi}} \frac{\partial  \ln \rho_F (\underline{\varphi},t) }{\partial \varphi(x_i)}
 \, , \label{eqn:eq-mom_1}
\end{eqnarray}
where, for the sake of later convenience, we reexpressed the right-hand side by using the following relations:
\begin{eqnarray}
v^2 (x_{i+1}, \underline{\varphi},t)  - v^2 (x_{i-1}, \underline{\varphi},t) 
&=& 
2 v (x_{i}, \underline{\varphi},t) \{ v (x_{i+1}, \underline{\varphi},t)  - v (x_{i-1}, \underline{\varphi},t) \} \, , \\
\sum_{i=-N}^{N-1} \partial_x V(\varphi(x_i) )
&=&
\sum_{i=-N}^{N-1}  (\partial_x \varphi (x_i)) \frac{\partial}{\partial \varphi(x_i)} V(\varphi(x_i)) \, . 
\end{eqnarray}
These are obtained by assuming that the difference between $\varphi(x_{i+1})$ ($v (x_{i+1}, \underline{\varphi},t)$) 
and $\varphi(x_{i})$ ($v (x_{i}, \underline{\varphi},t)$) becomes infinitesimal in the continuum limit.
One can show that the spatial integral of the right-hand side of Eq.\ (\ref{eqn:eq-mom_1}) 
vanishes and thus the momentum is conserved, because
\begin{eqnarray}
\sum_{i=-N}^{N-1} \sum_{j=-N}^{N-1} A \frac{\partial^2 B}{\partial \varphi (x_i) \partial \varphi(x_j)} \frac{\partial}{\partial \varphi(x_j)} \partial_x \varphi(x_i)
&=&
\frac{1}{2\Delta x}\sum_{i=-N}^{N-1} \sum_{j=-N}^{N-1} A \frac{\partial^2 B}{\partial \varphi (x_i) \partial \varphi(x_j)} 
\left( 
\delta_{j,i+1} - \delta_{j,i-1}
\right) \nonumber \\
&=& 0 \, , \label{eqn:formula1}
\end{eqnarray}
where $A$ and $B$ are functionals which are not explicitly depend on $x_i$ such as $ \rho_F (\underline{\varphi},t)$.

Note that the change of the momentum is characterized  by the gradient of the thermodynamical pressure in standard hydrodynamics, 
but the functional pressure $P_F$ vanishes in Eq.\ (\ref{eqn:eq-mom_1}).
As a matter of fact, $P_F$ is not the quantity comparable to the thermodynamical pressure in the spatial configuration space. 
In scalar ideal hydrodynamics, the canonical pressure is induced from the remaining field degrees of freedom in Eq.\ (\ref{eqn:eq-mom_1}) 
and will correspond to the thermodynamical pressure.
To see this, we utilize a procedure similar to the derivation of hydrodynamics from the Boltzmann equation \cite{hydro_review,boltzmann}.
We expand $\rho_F (\underline{\varphi},t)$ around a local equilibrium distribution in the function space as  
\begin{eqnarray}
\rho_F (\underline{\varphi},t)= \rho_{eq} (\underline{\varphi},t) + \delta \rho_F (\underline{\varphi},t) \, .
\end{eqnarray}
This decomposition satisfies
\begin{eqnarray}
\int [\ud \underline{\varphi}] \, \delta \rho_F (\underline{\varphi},t) = 0 \, .
\end{eqnarray}
Assuming further that the deviation $\delta \rho_F$ is small and higher order terms are negligible,  
Eq.\ (\ref{eqn:eq-mom_1}) is reexpressed as 
\begin{eqnarray}
\partial_t \langle p (x_i, t) \rangle_F 
&\approx& 
- \partial_x  \left\{\frac{1}{2c^2} u^2 (x_i,t) +  P_{can} (x_i,t)  \right\}\nonumber \\
&& -  c^{-2}  \frac{\nu^2}{ (\Delta x)^2} \int^\infty_{-\infty}  [\ud \underline{\varphi}] \,   (\partial_x \varphi (x_i))  
\partial_{\underline{\varphi}} * \left\{ \rho_F (\underline{\varphi},t) 
\partial_{\underline{\varphi}} \frac{\partial  \ln \rho_F (\underline{\varphi},t) }{\partial \varphi(x_i)}
\right\}
 \, .\label{eqn:eq-mom_2}
\end{eqnarray}
where
\begin{eqnarray}
u (x_i,t) = \int [\ud \underline{\varphi}] \, \rho_{eq} (\underline{\varphi},t) v(x_i, \underline{\varphi},t)  \, ,
\end{eqnarray}
and 
the canonical pressure is defined by  
\begin{eqnarray}
P_{can} (x,t)
=\int [\ud \varphi] \, \rho_{eq} (\underline{\varphi},t) \left[ 
\frac{1}{2c^2} (v (x,\underline{\varphi},t) - u (x,t) )^2 + \frac{1}{2} (\partial_x \varphi(x))^2 - V (\varphi(x))
 \right] \, . \label{eqn:pcan}
\end{eqnarray}
In fact, this quantity is reproduced from the canonical energy-momentum tensor $T^{ii} (\underline{\varphi})$, 
\begin{eqnarray}
P_{can} (x,t)
&=& \left.\frac{1}{D}  \int [\ud \varphi] \, \rho_{eq} (\underline{\varphi},t) \sum_{i=1}^D  T^{ii} (\underline{\varphi})  \right|_{D=1}\, ,
\end{eqnarray}
using the classical Lagrangian,  
\begin{eqnarray}
L 
= \int \ud x\, \left\{ \frac{1}{2c^2}(\partial_t \phi(x,t) - u (x,t) )^2 - \frac{1}{2} (\partial_x \phi(x,t))^2 - V (\phi (x))   \right\}  \, .
\end{eqnarray}
The appearance of $u (x,t)$ will be associated with the fact that thermal equilibrium is applied in the co-moving frame of the functional fluid element.

Let us ignore the last term on the right-hand side of Eq.\ (\ref{eqn:eq-mom_2}), which represents the correction to the pressure induced by fluctuation. 
We then find that Eq.\ (\ref{eqn:eq-mom_2}) has a similar structure to the momentum equation of ideal hydrodynamics in identifying 
the canonical pressure with the thermodynamical pressure.
Of course, to establish this correspondence relation more precisely, we have to show, for the relativistic case \cite{hugo},  
\begin{eqnarray}
\langle p (x_i, t) \rangle_F  &\longrightarrow& ( \varepsilon + P ) \gamma^2 \mathfrak{v} \, ,\\
\frac{1}{2c^2} u^2 (x_i,t) &\longrightarrow& (\varepsilon + P )\gamma^2 \mathfrak{v}^2 \, ,
\end{eqnarray}
or, for the non-relativistic case
\footnote{In the present case, the field Lagrangian has no conserved charge and thus it is not obvious whether the corresponding non-relativistic limti exists or not. 
See also Ref.\ \cite{gustavo}. }
,  
\begin{eqnarray}
\langle p (x_i, t) \rangle_F  &\longrightarrow& \rho_{mas} \mathfrak{v} \, ,\\
\frac{1}{2c^2} u^2 (x_i,t) &\longrightarrow& \rho_{mas} \mathfrak{v}^2 \, .
\end{eqnarray}
Here  $\mathfrak{v}$ is the fluid velocity in the configuration space,  $\gamma = 1/\sqrt{1 -(\mathfrak{v}/c)^2}$ and $\rho_{mas}$ denotes the mass density.
Although the correspondence relation has not yet been established,  
it is clear that a further coarse-graining is needed to obtain standard hydrodynamics from scalar ideal hydrodynamics.
This means that scalar ideal hydrodynamics is regarded as the mesoscopic theory such as the Boltzmann equation 
in the dynamical hierarchy of many-body systems.

\section{functional Schr\"{o}dinger equations} \label{sec:nonlinearSEQ}

Scalar ideal hydrodynamics defined by Eqs.\ (\ref{eqn:fih-1}), (\ref{eqn:fih-2}) and (\ref{eqn:fih-3}) is not manifestly Lorentz covariant 
and thus one may consider that the present model is applicable to describe only non-relativistic fluids.
Indeed, as was discussed,  the structures of these equations seem to be the functional generalization of non-relativistic ideal hydrodynamics.  
Contrary to this intuition, however, scalar ideal hydrodynamics reproduces the exact behavior of relativistic quantum field theory 
in a certain limit.

To see this, we need to reexpress scalar ideal hydrodynamics in a more familiar form. 
We first introduce the phase functional $\theta (\underline{\varphi},t)$ so as to satisfy  
\begin{eqnarray}
v(x_i,\underline{\varphi},t) = \frac{2\nu}{\Delta x} \frac{\partial}{\partial \varphi(x_i)} \theta (\underline{\varphi},t) \, .
\end{eqnarray}
We further consider that the barotropic equation of state where 
the functional pressure is a functional only of the functional probability distribution $\rho_F (\underline{\varphi},t)$.
In this case, we find the following reexpression: 
\begin{eqnarray}
\frac{1}{\rho_F (\underline{\varphi},t)} \frac{\partial}{\partial \varphi(x_i)} P_F (\rho_F)
= \frac{\partial}{\partial \varphi(x_i)} \int^{P_F(\underline{\varphi},t)} \frac{\ud \tilde{P}}{\rho_F (\tilde{P})} \, .
\end{eqnarray}
Then Eqs.\ (\ref{eqn:fih-1}) and (\ref{eqn:fih-3}) are reduced into the unified equation,
\begin{eqnarray}
\ii \partial_t \Psi  (\underline{\varphi},t)
&=& 
c^2 \left[- \frac{\nu}{c^2 \Delta x} \partial_{\underline{\varphi}}^2
-  \frac{\Delta x}{4\nu} \underline{\varphi} * \partial^2_x \underline{\varphi} 
+ \frac{\Delta x}{2\nu} \sum^{N-1}_{i=-N} V 
+ \frac{1}{2\nu}  \int^{P_{F}(\underline{\varphi},t)} \frac{\ud \tilde{P}}{\rho_F (\tilde{P})}
\right] \Psi  (\underline{\varphi},t)\, . \label{eqn:prot-fsc} \nonumber \\
\end{eqnarray}
where the wave functional is defined by 
\begin{eqnarray}
\Psi (\underline{\varphi},t) = \sqrt{\rho_F (\underline{\varphi},t) } e^{\ii \theta (\underline{\varphi},t) } \, .
\end{eqnarray}
See also the discussion in Ref.\ \cite{svm-field}.

It is easy to notice that Eq.\ (\ref{eqn:prot-fsc}) coincides with the functional Schr\"{o}dinger equation for the real scalar field 
by ignoring the last term associated with the functional pressure on the right-hand side,
\begin{eqnarray}
\ii \hbar \partial_t \Psi  (\underline{\varphi},t)
= 
H \Psi (\underline{\varphi},t)  \, , \label{eqn:fsch}
\end{eqnarray}
where the Hamiltonian operator is defined by 
\begin{eqnarray}
H = \int^{\infty}_{-\infty} \ud x\, \left( - \frac{\hbar^2 c^2}{2}  \frac{\delta^2}{\delta \varphi^2 (x) }  
- \frac{1}{2} \varphi(x) \partial^2_x \varphi(x) 
+  V(\varphi (x))
\right) \, ,
\end{eqnarray}
and we chose $\nu =\hbar c^2/2$.
The functional Schr\"{o}dinger equation (\ref{eqn:fsch}) reproduces the exact behaviors of relativistic quantum field theory 
and thus is consistent with relativistic kinematics \cite{svm-field,jackiw}.
Therefore,  
scalar ideal hydrodynamics is applicable to describe a relativistic ideal fluid unless the functional pressure term violates relativistic kinematics. 
It is also worth mentioning that there is an attempt to express the functional Schr\"{o}dinger equation in the manifestly coordinate-covariant form \cite{freese}.

When the functional pressure term is sustained, Eq.\ (\ref{eqn:prot-fsc}) becomes a non-linear equation.
For example, choosing $\varepsilon_F = \alpha \rho_F^2$ with $\alpha$ being a constant, 
the functional Schr\"{o}dinger equation is reexpressed as
\begin{eqnarray}
\ii \hbar \partial_t \Psi  (\underline{\varphi},t)
= 
\left[
 \int^{\infty}_{-\infty} \ud x\, \left( - \frac{\hbar^2 c^2}{2}  \frac{\delta^2}{\delta \varphi^2 (x) }  
- \frac{1}{2} \varphi(x) \partial^2_x \varphi(x) 
+  V(\varphi (x))
\right)
+ 4 \alpha |\Psi (\underline{\varphi},t)|^2
\right]
 \Psi (\underline{\varphi},t)  \, . \nonumber \\
\end{eqnarray}
The appearance of the non-linear term here is reminiscent of a similar behavior in quantum mechanics: 
the Sch\"{o}dinger equation is modified to the Gross-Pitaevskii equation when 
the internal energy term is added to the many-body Lagrangian \cite{koide12,review_ucr}.

As another hydrodynamical aspect of relativistic quantum field theory, 
the time-independent functional Sch\"{o}dinger equation is equivalent to the functional generalization of the Bernoulli equation.
See the discussion in Appendix \ref{app:bernoulli}.

\section{Summary and Concluding remarks} \label{sec:conc}

We studied new hydrodynamics to describe a fluid composed of the $1+1$ dimensional real scalar field.
This fluid is observed in the coarse-grained scale where the rapid oscillating part of the scalar field are already thermalized in the function space. 
The motion of this fluid is characterized by the trajectory of the scalar-field configuration.
Because of coarse-graining and fluctuation, these trajectories are distributed in the function space with a certain probability. 
We divided these trajectories into the bundles of fluxes, which correspond to the motions of the fluid elements (the functional fluid elements), 
and assumed that its internal state is thermally equilibrated. 
Then ideal hydrodynamics for the effective scalar field is derived by applying  
the stochastic variational method.
The equations of scalar ideal hydrodynamics are summarized by Eqs.\ (\ref{eqn:fih-1}), (\ref{eqn:fih-2}) and (\ref{eqn:fih-3}).
The first two equations represent the conservation laws of the probability and the energy, respectively.
The last equation describes the acceleration of the fluid in the function space. 
To obtain hydrodynamics in the spatial configuration space, scalar ideal hydrodynamics should be coarse-grained.
That is, scalar ideal hydrodynamics is regarded as the mesoscopic theory such as the Boltzmann equation in the dynamical hierarchy of many-body systems.

The functional local equilibrium ansatz requires that 
thermodynamics is satisfied in the local domain of the function space and thus is not equivalent to the local equilibrium ansatz in the spatial configuration space.  
One may thus doubt such an application of thermodynamics.
The framework of thermodynamics is however considered to be robust and has been applied even to abstract concepts like information: 
Maxwell's daemon, Szilard's engine, Landauer's principle and so on. See, for example, Ref.\ \cite{information} and references therein.
Our attempt developed in this paper is an extension of this line of study. 
Indeed, the assumptions of thermodynamics in the function space (Eqs.\ (\ref{eqn:ft}) and (\ref{eqn:ther_euler})) 
are consistent in the sense that
the equations of the energy and entropy distributions satisfy conservation laws.
See also the discussion for the thermodynamical relations in the function space in Ref.\ \cite{scalar-SE}.

In quantum mechanics, the evolution of the wave function is known to be expressed by the ensemble of the trajectories of quantum particles 
which is called the Bohmian trajectory \cite{book_holland}.
The trajectories are determined by solving the Newton equation with the quantum potential.
The deterministic motion of the functional fluid element is defined by the classical field equation with the functional quantum potential 
and thus the trajectory described by Eq.\ (\ref{eqn:tra}) can be interpreted as the functional generalization of the Bohmian trajectory.
It should be however noted that this trajectory is not directly related to 
the motion of a quantized particle. For the relation between the Bohmian trajectory and the particle trajectory in quantum field theory, 
See Ref.\ \cite{durr}.

Although the deterministic motion given by Eq.\ (\ref{eqn:tra}) is convenient to consider the Bohmian interpretation in quantum field theory,  
we consider that the more interesting trajectory of the functional fluid element is given by the (Nelson-type) zigzag paths described by 
SDE's (\ref{eqn:fsde}) and (\ref{eqn:bsde}).
It is already known in quantum mechanics that the well-known uncertainty relations 
(the Kennard and the Robertson-Schr\"{o}dinger inequalities) are reproduced by evaluating the uncertainty induced 
by the non-differentiability of the trajectory of a quantum particle \cite{review_ucr,koide_ucr1,koide_ucr2}. 
This method resolves the paradox of the angular uncertainty relation \cite{koide_ucr3} 
and predicts the uncertainty relations associated with viscosity \cite{koide_ucr1}.
The minimum uncertainty state of a viscous fluid is studied in Ref.\ \cite{koide_ucr2}.
These studies are limited to non-relativistic dynamics so far, but the present work enables us to generalize these ideas to relativistic systems.

In this paper, scalar ideal hydrodynamics is constructed using mathematical analogy between the spatial particle position and the field configuration in the function space. 
Our formulation is consistent mathematically, but we need a special attention to the physical interpretation.
In standard hydrodynamics, for example, when there is a spatial domain which has a higher pressure, the distribution of the fluid spreads out by diffusion 
which is caused by the intermolecular interaction among the constituent particles of the fluid. When a similar distribution is realized in the function space,
 we assumed that diffusion is induced by the interactions among different field configurations which constitute the functional fluid. 
This diffusion is however caused by a statistical force such as noise in Brownian motion, and this force is induced by fluctuations associated 
with quantum effects and coarse-graining.
The applicability of this physical perspective should be investigated carefully.

It is important to note that scalar ideal hydrodynamics is not expressed in a manifestly Lorentz-covariant form, 
but, may be applicable to describe relativistic behaviors.
Indeed, if we ignore the contribution from the functional pressure term, scalar ideal hydrodynamics is reduced to 
the functional Schr\"{o}dinger equation for the scalar field and reproduces the exact behaviors of relativistic quantum field theory. 
Therefore our theory may provide a new relativistic model of hydrodynamics unless the functional pressure term violates relativistic kinematics. 
Functional hydrodynamics is then useful to study the influence of quantum fluctuation in collective flows of 
the produced particles in relativistic heavy-ion collisions.

The following issues are left as future tasks.
In this work, functional hydrodynamics was formulated by assuming that the macroscopic behaviors are described 
by the coarse-grained real scalar field. 
However the macroscopic behaviors are not necessarily given by the scalar field. 
We should choose carefully fields to construct functional hydrodynamics.
The formulation developed in this paper will be applicable to general field theories like gauge fields.
The present field-theoretical approach to hydrodynamics is reminiscent of the studies of soliton stars where the fundamental degrees of freedom of, for example, 
the neutron star is described by fields and then the behavior of the star is described through solitons formed by the non-linear equation of the fields 
\cite{lee,nugaev,glendenning}.
Our hydrodynamics thus may be useful to investigate the effect of solitons in relativistic heavy-ion collisions and neutron star mergers.
We have considered the formulation of ideal hydrodynamics. In SVM, the effect of viscosity is automatically introduced by
considering more general quadratic forms in the replacement of the time derivative terms in the stochastic Lagrangian (\ref{eqn:sto-lag}).
See, for example, Refs.\ \cite{review_ucr,koide12}.
Applying this to Eq.\ (\ref{eqn:sto-lag}), we obtain scalar viscous hydrodynamics.

\vspace*{1cm}
T.\ Koide thanks to A.\ F. Costa for a fruitful discussion.
The authors acknowledge the financial supports by CNPq (No.\ 305654/2021-7, \, 303246/2019-7),
FAPERJ and CAPES. 
A part of this work has been done under the project INCT-Nuclear Physics and Applications (No.\ 464898/2014-5).

\appendix

\section{Variational derivation of pressure} \label{app:pressure}

The variation associated with the internal energy of the functional fluid element does not depend on the time derivatives 
and hence the operation of stochastic variation to this term is the same as that of the classical variation.
Therefore we consider only the classical variation in the following discussion.

The functional fluid element which is initially located at $\underline{\phi}_0$ moves to a new position $\underline{\phi}_{eff} (t)$ as time goes by.
This time evolution can be regarded as the coordinate transformation.
Therefore the probability distribution, which is normalized by one, is expressed as \cite{review_ucr}
\begin{eqnarray}
\rho_F (\underline{\phi}_{eff}(t),t) = \frac{1}{J(\underline{\phi}_{eff}(t), \underline{\phi}_0)} \rho_0 (\underline{\phi}_0) \, ,
\end{eqnarray}
where the Jacobian is given by 
\begin{eqnarray}
J (\underline{\phi}, \underline{\phi}_0) = \left| \frac{\partial \underline{\phi} }{\partial \underline{\phi}_0}\right| \, .
\end{eqnarray}
Then the variation of the probability distribution induced by Eq.\ (\ref{eqn:variation_phi}) is given by 
\begin{eqnarray}
\delta \rho_F (\underline{\phi}_{eff}(t), t)  
&=&
- \frac{\rho_F (\underline{\phi}_{eff}(t), t)  }{J (\underline{\phi}_{eff} (t),\underline{\phi}_0)} \delta J \nonumber \\
&=& 
- \frac{\rho_F (\underline{\phi}_{eff}(t),t)}{J (\underline{\phi}_{eff}(t),\underline{\phi}_0)}  \sum_{ij=0}^{N-1} A_{ij} \frac{\partial }{\partial \phi_0 (x_j)} \delta f
( x_i,\underline{\phi}_{eff} (t), t ) \, ,
\end{eqnarray}
where $A_{ij}$ are the cofactors associated with the Jacobian.
Note that the cofactor satisfies the following properties:
\begin{eqnarray}
\sum_{j=-N}^{N-1}\frac{\partial}{\partial \phi_0(x_j)} A_{ij} &=& 0 \,, \\
\sum_{j=-N}^{N-1} A_{ij} \frac{\partial}{\partial \phi_0(x_j)}
&=& 
\sum_{j=-N}^{N-1} A_{ij} \sum_{k=-N}^{N-1} \frac{\partial \phi_{eff} (x_k,t)}{\partial \phi_0 (x_j)} \frac{\partial}{\partial \phi_{eff}(x_k,t)} 
\nonumber \\
&=&  J (\underline{\phi}_{eff}(t),\underline{\phi}_0) \frac{\partial}{\partial \phi_{eff}(x_i,t)} \, .
\end{eqnarray}
Using these, it is straightforward to show the last line of Eq.\ (\ref{eqn:vari_int_ener}).

\section{``Kinematic" approach to the transport in the function space} \label{app:kinetic}

As discussed, $\underline{\phi}(t)$ corresponds to the position in the function space.
Generalizing this idea, we consider a kind of the phase space description of classical real scalar field theory.
The functional phase space of the real scalar field is defined by (2N $\times$2N) variables, 
$(\underline{\phi}(t), \underline{\Pi}(t))$ where $\underline{\Pi}(t) =(\Pi(x_{-N},t),\Pi(x_1,t),\cdots,\Pi(x_{N-1},t))$ is the conjugate field of $\underline{\phi}(t)$.
The time evolutions of the classical fields are described by the Klein-Gordon equation, 
\begin{eqnarray}
\partial_t \phi(x_i,t) &=& c^2 \Pi (x_i,t) \, ,\\
\partial_t \Pi(x_i,t) &=& \nabla^2 \phi(x_i,t) - \frac{\partial V (\phi(x_i,t))}{\partial \phi(x_i,t)} \, .
\end{eqnarray}
The motion of the classical fields is described by a continuous trajectory in the functional phase space which is determined by solving the Klein-Gordon equation.

Let us introduce the probability distribution of the trajectory, 
\begin{eqnarray}
f(\underline{\varphi},\underline{\varPi},t) = \prod_{i=-N}^{N-1} \delta(\varphi(x_i)- \phi(x_i,t)) \delta(\varPi(x_i)- \Pi(x_i,t)) \, .
\end{eqnarray}
Here we do not consider the initial distribution of $(\underline{\phi}, \underline{\Pi})$.
One can easily see that this is a generalization of $\rho_F (\underline{\varphi},t)$ in Eq.\ (\ref{eqn:rho_cla}).
See also the discussion in Ref.\ \cite{scalar-SE}.
Then the Liouville equation in the function space reads 
\begin{eqnarray}
\partial_t f(\underline{\varphi},\underline{\varPi},t) = - \sum_{i=-N}^{n-1} \left[
\frac{\partial}{\partial \varphi(x_i)} \varPi(x_i)
+ \frac{\partial}{\partial \varPi (x_i)} \left\{ \partial^2_x \varphi(x_i)  - \frac{\partial V (\phi(x_i,t))}{\partial \phi(x_i,t)}\right\}
\right] f(\underline{\varphi},\underline{\varPi},t) \, . \label{eqn:f_liouville}
\end{eqnarray}
This equation reproduces the exact behavior of the classical Klein-Gordon equation.

When we describe the system as a functional only of $\underline{\phi}$, we can introduce the following moments,
\begin{eqnarray}
\rho_{\varphi} (\underline{\varphi},t) &=& \int [\ud \underline{\varPi}] f(\underline{\varphi},\underline{\varPi},t) \, , \\
v_\varphi (x_i,\underline{\varphi},t) &=& \frac{1}{\rho_\varphi (\underline{\varphi},t)} \int [\ud \underline{\varPi}] \varPi(x_i)  f(\underline{\varphi},\underline{\varPi},t) \, .
\end{eqnarray}
The evolution equations for these moments can be derived from Eq.\ (\ref{eqn:f_liouville}).
It is easy to show that the probability distribution $\rho_{\varphi} (\underline{\varphi},t)$ satisfies the same equation of continuity as Eq.\ (\ref{eqn:feoc}). 
On the other hand, the equation for $v_\varphi (x_i,\underline{\varphi},t) $ is 
\begin{eqnarray}
\partial_t  v_\varphi(x_i, \underline{\varphi}, t) + \underline{v}_\varphi(\underline{\varphi}, t) * \partial_{\underline{\varphi}}  v_\varphi (x_i, \underline{\varphi}, t)
= -\frac{1}{\rho_\varphi (\underline{\varphi},t)} \sum_{j=-N}^{N-1} \frac{\partial}{\partial \varphi(x_i)} p_{ij} + \partial^2_x \varphi(x_i) - \frac{\partial V (\varphi(x_i))}{\partial \varphi(x_i)} \, , \nonumber \\
\end{eqnarray}
where
\begin{eqnarray}
p_{ij} = \int [\ud \underline{\varPi}] (\varPi(x_i)- v_\varphi(x_i,\underline{\varphi},t))(\varPi(x_j)- v_\varphi(x_j,\underline{\varphi},t)) f(\underline{\varphi},\underline{\varPi},t)  \, .
\end{eqnarray}
It is easy to find that we can identify ($\rho_{\varphi}( \underline{\varphi},t), v_\varphi (x_i,\underline{\varphi},t)$ with $(\rho_{F} (\underline{\varphi},t), v (x_i,\underline{\varphi},t))$ when we choose 
\begin{eqnarray}
p_{ij}
= 
\frac{c^2}{\Delta x}P^{ij}_F (\rho_F,S_F) \, ,
\end{eqnarray}
where $P^{ij}_F (\rho_F,S_F)$ is defined in Eq.\ (\ref{eqn:eff-pressure}).

The equation for the phase space distribution (\ref{eqn:f_liouville}) can be regarded as a kind of the Boltzmann equation without the collision term.
If we can introduce the collision term in Eq.\ (\ref{eqn:f_liouville}) so as to reproduce $P^{ij}_F (\rho_F,S_F)$, 
it might be possible to study relativistic quantum field theory form the kinematic point of view.

\section{functional Bernoulli equation} \label{app:bernoulli}

The time-independent functional Schr\"{o}dinger equation is expressed by 
\begin{eqnarray}
H \Psi  (\underline{\varphi},t)
= 
e \Psi (\underline{\varphi},t)  \, , \label{eqn:tinscheq}
\end{eqnarray}
where $e$ is a constant. 
This equation is equivalent to the conservation of the Bernoulli functional. 
From Eq.\ (\ref{eqn:fih-3}), we obtain 
\begin{eqnarray}
\frac{\Delta x}{2c^2} \partial_t \{ \underline{v}(\underline{\varphi},t) *\underline{v}(\underline{\varphi},t) \} 
= - \underline{v}(\underline{\varphi},t) * \partial_{\underline{\varphi}} B_{er} \, ,  \label{eqn:dif-ber}
\end{eqnarray}
where the Bernoulli functional is defined by 
\begin{eqnarray}
B_{er} 
&=& (\Delta x) \left[ \frac{1}{2c^2}  \underline{v}(\underline{\varphi},t) *\underline{v}(\underline{\varphi},t) 
- \frac{1}{2} \underline{\varphi} * \partial^2_x \underline{\varphi} + \sum_{i=-N}^{N-1} V(\varphi(x_i)) \right. \nonumber \\
&& \left. + \frac{1}{\Delta x}  \int^{P_F} \frac{\ud \tilde{P}}{\rho_F (\tilde{P})}  
-\frac{2\nu^2}{c^2 (\Delta x)^2} \frac{1}{\sqrt{\rho_F}} \partial^2_{\underline{\varphi}} \sqrt{\rho_F} \right] \, .
\end{eqnarray}
Here we considered the barotropic equation of state.

Let us consider a stationary fluid, where the left-hand side of Eq.\ (\ref{eqn:dif-ber}) vanishes. 
Moreover, $\underline{v}(\underline{\varphi},t) * \partial_{\underline{\varphi}}$ represents the differential along the flow of the functional fluid element. 
Thus $B_{er}$ is constant for the stationary fluid,
\begin{eqnarray}
e
&=&
(\Delta x) \left[ 
\frac{2\nu^2}{(\Delta x)^2c^2}  
\left[
\partial_{\underline{\varphi}} \theta* \partial_{\underline{\varphi}} \theta 
- \frac{1}{\sqrt{\rho_F}} \partial^2_{\underline{\varphi}} \sqrt{\rho_F} 
\right]
- \frac{1}{2} \underline{\varphi} * \partial^2_x \underline{\varphi}  + \sum_{i=-N}^{N-1} V(\varphi(x_i)) 
+ \frac{1}{\Delta x}  \int^{P_F} \frac{\ud \tilde{P}}{\rho_F (\tilde{P})} \right] \, . \nonumber \\
\end{eqnarray}
This conservation law is the functional generalization of the Bernoulli theorem in hydrodynamics, and  
it is straightforward to confirm that this is equivalent to the time-independent functional Schr\"{o}dinger equation (\ref{eqn:tinscheq}).
See also Appendix B in Ref.\ \cite{review_ucr}.

\end{document}